\documentclass{ifacconf}

\usepackage{natbib}

\usepackage{amsmath} 
\usepackage{amssymb}  
\usepackage{amsfonts}  
\usepackage{graphics}
\usepackage{pdflscape}
\usepackage{rotating}
\usepackage{epstopdf}
\usepackage{subcaption}

\def\bul{\noindent $\bullet\;\;$}

\newcommand{\col}{\mbox{col}}

\def\L2e{{\cal L}_{2e}}
\def\bul{\noindent $\bullet\;\;$}
\def\rea{\mathbb{R}}

\def\adj{\mbox{adj}}
\def\liminf{\lim_{t \to \infty}}

\def\begequarr{\begin{eqnarray}}
\def\endequarr{\end{eqnarray}}
\def\begequarrs{\begin{eqnarray*}}
\def\endequarrs{\end{eqnarray*}}
\def\begarr{\begin{array}}
\def\endarr{\end{array}}
\def\begequ{\begin{equation}}
\def\endequ{\end{equation}}
\def\lab{\label}
\def\begdes{\begin{description}}
\def\enddes{\end{description}}
\def\begenu{\begin{enumerate}}
\def\begite{\begin{itemize}}
\def\endite{\end{itemize}}
\def\endenu{\end{enumerate}}
\def\lef[{\left[\begin{array}}
\def\rig]{\end{array}\right]}

\def\begcen{\begin{center}}
\def\endcen{\end{center}}

\def\calh{\mathcal{H}}

\def\caly{\mathcal{Y}}

\def\call{\mathcal{L}}
\def\calz{\mathcal{Z}}

\def\epst{\epsilon_t}
\def\et{\epsilon_t}

\def\rea{\mathbb{R}}

\usepackage{color}

\def\begmat#1{\begin{bmatrix}#1\end{bmatrix}}
\def\begali#1{\begin{align}{#1}\end{align}}
\def\begalis#1{\begin{align*}{#1}\end{align*}}

\usepackage[prependcaption,colorinlistoftodos]{todonotes}

\graphicspath{{}{fig/}}

\begin{document}

\begin{frontmatter}

\title{Sensorless Control of the Levitated Ball\thanksref{footnoteinfo}}

\thanks[footnoteinfo]{Corresponding author A.~A.~Vedyakov.}

\author[ITMO]{Alexey Bobtsov}
\author[ITMO]{Anton Pyrkin}
\author[SPLC]{Romeo Ortega}
\author[ITMO]{Alexey Vedyakov}

\address[ITMO]{School of Computer Technologies and Control, ITMO University, Kronverksky av., 49, 197101, Saint Petersburg, Russia (e-mail: bobtsov@mail.ru, pyrkin@corp.ifmo.ru, vedyakov@corp.ifmo.ru)}
\address[SPLC]{Laboratoire des Signaux et Systèmes, CNRS-SUPELEC, Plateau du Moulon, 91192, Gif-sur-Yvette, France (e-mail: ortega@lss.supelec.fr)}

										
\begin{abstract}
One of the most widely studied dynamical systems in nonlinear control theory is the levitated ball. Several full-state feedback controllers that ensure asymptotic regulation of the ball position have been reported in the literature. However, to the best of our knowledge, the design of a stabilizing law measuring only the current and the voltage---so-called sensorless control---is conspicuous by its absence. Besides its unquestionable theoretical interest, the high cost and poor reliability of position sensors for magnetic levitated systems, makes the problem of great practical application. Our main contribution is to provide the fist solution to this problem.  Instrumental for the development of the theory is the use of parameter estimation-based observers, which combined with the dynamic regressor extension and mixing parameter estimation technique, allow the reconstruction of  the magnetic flux. With the knowledge of the latter it is shown that the mechanical coordinates can be estimated with suitably tailored nonlinear observers. Replacing the observed states, in a certainty equivalent manner, with a full information asymptotically stabilising law completes the sensorless controller design. Simulation results are used to illustrate the performance of the proposed scheme.
\end{abstract}

\end{frontmatter}

%
\section{Introduction}
\lab{sec1}
%
Because of the poor observability properties of magnetic levitation systems, the problem of controlling their position assuming that only the current and the voltage are measurable---that is, the so-called sensorless (or self-sensing) scenario---is theoretically very challenging. Moreover, the high cost and low reliability of existing position sensors makes the problem practically important. For the latter reason, a lot of research has been devoted to the development of technologically-based techniques for sensorless control by the applications community \cite{RANetal,schw09book}. On the other hand, theoretically-based designs of state observers proceeding from the mathematical model of the system have also been reported by the control community \cite{GLUetal,MASetalifac,MIZetal}. As is well-known, the dynamic behavior of these systems is highly nonlinear. Therefore, to ensure good performance in a wide operating range it is necessary to avoid the use of linearized models that, to the best of the authors' knowledge, is the prevailing approach reported in the literature \cite{GLUetal,MIZetal}. See \cite{MASetal,MON} for a detailed analysis of the deleterious implications of linearization in sensorless Maglev models. 

In this paper we address the problem of sensorless control of the levitated ball system. Unquestionably, this is one of the most widely studied systems in the control community, with many educational labs disposing of experimental facilities for them. Although many full-state feedback asymptotically stabilizing controllers are available in the literature, see {\em e.g.}, \cite{BONetal,LEVetal,LINKNO,MASetalifac,ORTbook,TORORT}, to the best of our knowledge, no sensorless solution for the full nonlinear model has been reported. A notable exception is \cite{yicdc18} where the {\em signal injection} technique proposed in \cite{COMetal,yiscl18}, is used to give a solution to this problem.  The invasive injection of probing signals, that unavoidably degrades the transient performance, is avoided in the present contribution. On the other hand, as always for observer based controller designs for nonlinear systems, some excitation condition needs to be imposed on the signals of the system \cite{ARAetaltac}.

The present paper follows the same lines as the work for the two-degrees-of-freedom system reported in \cite{BOBetalaut18}. However, as shown below, the solution for the levitated ball turns out to be much more complicated. The first step in our design is the reconstruction of the flux, which is done by combining the parameter estimation-based observers (PEBO) recently reported in \cite{ORTetalscl} with the dynamic regressor extension and mixing (DREM) parameter estimation technique of \cite{ARAetaltac}---see also \cite{ORTetalaut} for the reformulation of DREM as a functional Luenberger observer.  With the knowledge of the flux we propose suitably tailored nonlinear observers for the mechanical coordinates, obtaining in this way a globally convergent solution to the posed observation problem. To complete the sensorless controller design the observed state is then replaced in the globally asymptotically stabilizing full-state feedback-linearizing controller (FLC) reported in \cite{ORTbook}. 

The remainder of the paper is organized as follows. Section \ref{sec2} briefly introduces the model of the levitated ball and formulates its state observer and sensorless control problems. Section \ref{sec3} presents the state observer. In Section \ref{sec4} the sensorless controller is presented. Simulation results are given in Section  \ref{sec5}. The paper is wrapped-up with concluding remarks and future research directions in \ref{sec6}.\\

\textbf{Notation.}
$|\cdot|$ is the Euclidean norm. $\et$ is a generic exponentially decaying term. For an operator $\calh$ acting on a signal we use the notation $\calh[\cdot](t)$, when clear from the context, the argument $t$ is omitted.

\section{Model and Problem Formulation}
\label{sec2}

The classical model of the unsaturated, levitated ball depicted in Fig. 1 is given as \cite{schw09book}
\begequ
\label{sys}
\begin{aligned}
    \dot{\lambda} & = - Ri + u \\
    \dot{Y} & = {1 \over m} p \\
    \dot{p} & = {1 \over 2k} \lambda^2 - mg\\
    \lambda &={k \over c-Y}i,
\end{aligned}
\endequ
where $\lambda$ is the flux linkage, $i$ the current, $Y \in (-\infty, c)$ is the position of the ball, $p$ is the momenta, $u$ is the input voltage, $R>0$ is the  resistance, and $m>0$, $c>0$ and $k>0$ are some constant parameters.

\begin{figure}
    \centering
    \includegraphics[width=3cm]{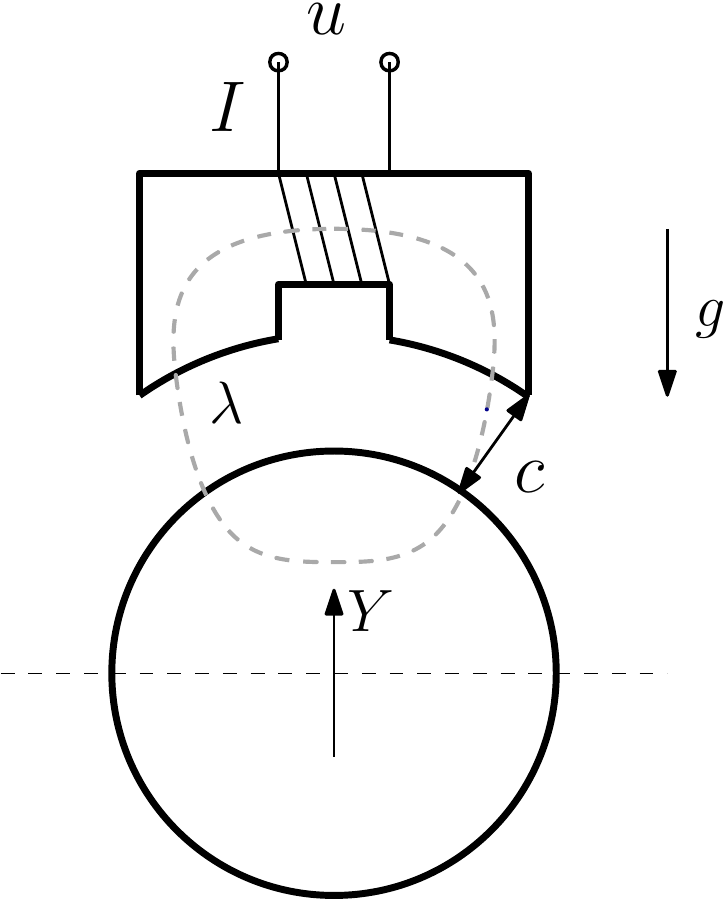}
    \label{fig:maglec}
    \caption{Schematic diagram of the levitated ball}
\end{figure}

In this paper we provide a solution to the following.\\

\noindent {\bf State Observer Problem.} Consider the dynamics of the levitated ball \eqref{sys}, with the parameters $m$, $c$, $k$ and $R$ known. Define the state vector \begequ
\lab{x}
x:=\col(Y,p,\lambda).
\endequ 
Design an observer
\begequ
\lab{adaobs}
\begin{aligned}
    \dot{\chi} & = F(\chi,u,i)\\
    \hat x & = H(\chi,u,i)
\end{aligned}
\endequ
where $\chi \in \rea^{n_{\chi}}$ is the observer state, such that
\begequ
\label{x_conv}
\limsup_{t\to\infty} |\hat x(t) - x(t)| =0.
\endequ

As usual in observer design problems we need the following.

\noindent {\bf Assumption A1}
Consider the system \eqref{sys}. The input signal $u$ is such that the state $x$ is {\em bounded}.

The sensorless controller is obtained applying certainty equivalence to the linear, static-state feedback, asymptotically stabilizing, FLC reported in \cite{ORTbook}, to ensure
\begequ
\lab{tracla}
 \limsup_{t \to \infty}  |Y(t) - Y_\star|=0,
\endequ
where $Y_\star$ is the desired position for the levitated ball.

\begin{rem}
\lab{rem1}
We make the important observation that it is possible to show that the system does not satisfy the observability rank condition [Section 1.2.1]\cite{bes07book}, therefore it is not uniformly differentially observable.
\end{rem}
%
\section{State Observer of the $1$-dof MagLev System}
\lab{sec3}
%
The observer is derived in five steps, which are treated in separate subsections.  

\subsection{Regression model for the PEBO of the flux}
\lab{subsec31}
%
The first, step for the observer design is to propose a PEBO for the flux of the form
\begequ
\lab{dotpsi0}
\dot \psi  =  -Ri+u.
\endequ 
From \eqref{sys} and \eqref{dotpsi0} we conclude that 
\begin{align}
\lab{x3_par}
\lambda(t)=\psi(t)+\eta,
\end{align}
where $\eta:=\lambda(0)-\psi(0)$. Following the PEBO design the problem is to estimate the parameter $\eta$, and reconstruct the flux from \eqref{x3_par}. Towards this end, it is necessary to establish a (nonlinear) regression for $\eta$, that is, an algebraic relation that depends only on the signals $y$ and $u$ and a function of the unknown parameter $\eta$---a result which is contained in the proposition below. Since the computations are pretty cumbersome, its proof is given in the Appendix.
 
\begin{prop}
\lab{pro1}
Consider the model of the $1$-dof Maglev system \eqref{sys} and the dynamic extension \eqref{dotpsi0}. The  constant parameter $\eta$ satisfies the following (nonlinearly parameterised) regression model
\begequ
\lab{reg9}
z = \phi^\top \Omega(\eta),
\endequ
where $z \in \rea$ and $\phi \in \rea^5$ are measurable signals and 
\begequ
\lab{regmodcom}
\Omega(\eta):=\col(\eta,\eta^2,\eta^3,\eta^4,\eta^5).
\endequ
\end{prop}

\begin{rem}
\lab{rem2}
The regression model  \eqref{reg9} is  {\em nonlinearly parameterised}. Although it is possible to obtain a linear regression introducing an overparameterisation, we avoid this low performance approach here. Instead,  we use DREM to estimate directly the parameter $\eta$ with just one gradient search.
\end{rem}

\begin{rem}
\lab{rem20}
Besides the additional difficulty of needing to estimate $\eta$, the main drawback of PEBO is that it relies on the open-loop integration \eqref{dotpsi0}, which might be a problematic operation in practice. For a discussion on this matter see \cite{MASIWA} where the open-loop integration \eqref{dotpsi0} is proposed---but without the essential parameter estimation step.
\end{rem}
\subsection{Parameter estimation via DREM}
\lab{subsec32}
%
Before presenting the flux DREM estimator we recall the following lemma, which will be instrumental in the proof of the main result.

\begin{lem}\cite{ARAetal}
\lab{lem1}
Consider the scalar, linear time-varying, system defined by $\dot{x} = -a^2(t) x + b(t)$,
where $x\in \mathbb{R}$, $a(t)$ and $b(t)$ are piecewise continuous functions. If $a(t) \not \in {\mathcal L}_2$ and $b(t) \in {\mathcal L}_1$ then $\lim\limits_{t \to \infty}{x(t)} = 0$.
\end{lem}

\begin{prop}
\lab{pro2}
Consider the model of the $1$-dof Maglev system \eqref{sys} with the regression model \eqref{reg9}. Fix four stable filters ${\kappa_j\over p+\nu_j}$, $j=1,\dots,4$, with $p:={d \over dt}$ and $\kappa_j>0,\nu_j>0$. Define the filtered signals
\begequ
\lab{deffil}
(\cdot)^{f_j}:={\kappa_j\over p+\nu_j}\left(\cdot\right),\;j=1,\dots,4,
\endequ
and generate the  DREM parameter estimates as
\begequ
\label{decesti0}
\dot{\hat{\eta}}  =  \gamma\Delta (\caly - \Delta \hat\eta),
\endequ
with gain $\gamma>0$, where we introduced the signals
\begali{
\lab{calz2}
\calz &:=  \col( z, z^{f_1}, \cdots,  z^{f_4}),\; \Phi^T :=  \begmat{ \phi & |& \phi^{f_1} & |&\cdots &|& \phi^{f_4}} \\
\label{caly}
\caly & :=  e_1^\top\adj\{\Phi\} \calz,\quad \Delta  :=  \det \{\Phi\},
}
where $e_1:=\col(1,0,0,0,0)$ and $\adj\{\cdot\}$ is the adjunct matrix.  Generate the flux estimate as
\begequ
\lab{fluest0}
\hat \lambda :=\psi + \hat \eta.
\endequ
The following implication is true
\begequ
\lab{imp0}
\Delta(t) \notin \call_2 \; \Rightarrow \; \liminf e_\lambda(t)=0.
\endequ
where we defined the flux estimation error $e_\lambda:=\hat \lambda - \lambda$.
\end{prop}

\begin{pf}
Applying the filters to the regressor model \eqref{reg9}, \eqref{regmodcom} and arranging terms we get 
$$
\calz=\Phi  \Omega(\eta).
$$
Premultiplying this by the {adjunct} of $\Phi$  and retaining the first {scalar} regressor we get $\caly=\eta\Delta$. Replacing the latter in \eqref{decesti0}, and using \eqref{x3_par} and \eqref{fluest0}, we get the flux error equation
\begequ
\lab{dotelam}
\dot{e}_\lambda = -\gamma \Delta^2 e_\lambda+\et.
\endequ
We complete the proof invoking Lemma \ref{lem1}.
\end{pf}
\subsection{Speed observer }
\lab{subsec33}
%
To prove convergence of the proposed speed observer we make the following, practically reasonable, assumption.

\noindent {\bf Assumption A2} The control voltages $u$ and the vertical speed $\dot Y$ are bounded.

Recalling that $\dot\lambda$ is measurable, we propose the following speed observer.

\begin{prop}
\lab{pro3}
Consider the model of the levitated ball system \eqref{sys} and the speed observer
\begin{align}
	\nonumber
	\dot\chi &= {1\over m} \left(
		{1\over 2k}\hat \lambda^2 -mg
	\right) - \gamma_v \hat \lambda^2 \hat v + 2 \gamma_v k i \dot\lambda,\\
	\lab{v_hat}
	\hat v & = \chi - \gamma_v k i \hat \lambda,
\end{align}
where $\gamma_v>0$, and $\hat \lambda$ is generated as in Proposition \ref{pro2}. The following implication is true
\begin{align}
\Delta\not\in \mathcal{L}_2 \;\mbox{and}\;\hat \lambda\not\in \mathcal{L}_2 \; \Rightarrow \; \lim_{t\rightarrow\infty} e_v(t)=0,
\lab{equspe0}
\end{align}
where we defined the speed estimation error $e_v:=\hat v - \dot Y$.
\end{prop}

\begin{pf}
Differentiating the last equation in \eqref{sys} and multiplying by $\lambda$ we get
$$
	k {di \over dt} \lambda - k i \dot\lambda =  - \dot Y \lambda^2,
$$
Using this and  the speed observer \eqref{v_hat} we get, after some simple manipulations, the error model
\begequ
\lab{dotev}
	\dot e_v = -\gamma_v \hat \lambda^2 e_v + \delta_v,
\endequ
where 
\begin{align}
\nonumber
\delta_v:= & \frac{1}{2km}\left(2\hat{\lambda}-e_{\lambda}\right)e_{\lambda}
-\gamma_{\upsilon}\hat{\lambda}\left(Y-c\right)\dot{e}_{\lambda}
\\
\nonumber
& 
-\gamma_{\upsilon}\left(k\frac{di}{dt}+2\hat{\lambda}\upsilon-\dot{e}_{\lambda}\left(Y-c\right)-e_{\lambda}\upsilon\right)e_{\lambda}.
\end{align}
The proof is completed noting that $\delta_v(t) \to 0$ and integrating the scalar equation above.
\end{pf}
\vspace{-2mm}
\subsection{Position observer}
\lab{subsec34}
%
The final step in the observer design is to reconstruct the position $Y$.

\begin{prop}
\lab{pro4}
Consider the model of the $1$-dof Maglev system \eqref{sys}. Define the position observer
\begequ
\lab{obsy0}
\dot{\hat Y}= - \gamma_Y \hat \lambda^2 \hat Y + \gamma_Y  (c\hat \lambda - ki)\hat \lambda + \hat v,
\endequ
where $\gamma_Y>0$, $\hat \lambda$ and $\hat v$ are generated as in Propositions \ref{pro2} and \ref{pro3}, respectively. The following implication is true
$$
\Delta\not\in \mathcal{L}_2 \;\mbox{and}\;\hat \lambda\not\in \mathcal{L}_2 \; \Rightarrow \;  \lim_{t\rightarrow\infty} e_Y(t)=0.
$$
where $e_Y:=\hat Y - Y$.
\end{prop}

\begin{pf}
Multiplying by $\lambda$ the last equation in \eqref{sys} we get 
$$
 (c\lambda - ki)\lambda=\lambda^2 Y
$$
which replaced in \eqref{obsy0}  yields 
\begequ
\lab{dotey}
\dot e_Y=-\gamma_Y \hat \lambda^2 e_Y  + \delta_Y,
\endequ
where 
$$
\delta_Y:=\gamma_{Y}\hat{\lambda}(c-Y)e_{\lambda} + e_{\upsilon}.
$$
The proof is completed noting that $\delta_Y(t) \to 0$. 
\end{pf}
%
\section{Sensorless Controller}
\lab{sec4}
%
In this section we implement the sensorless controller replacing, in a certainty equivalent way, the estimated flux, position and velocity described in the previous section, in the following FLC:
\begin{align}
\nonumber
u & = {k \over \lambda}mv_{FL} + R(c-Y)\frac{\lambda}{k}, \\
\nonumber
v_{FL} & = Y_\star ^{(3)} - k_2((\frac{\lambda^2}{2km} - g) - \ddot{Y}_\star )  \\
	& \quad - k_1(\dot{Y} - \dot{Y}_\star ) - k_0(Y - Y_\star ).
\lab{feelin}
\end{align}
This controller is given in Chapter 8, Section 5.1 of \cite{ORTbook}---see also \cite{LINKNO,TORORT}---and, replaced in \eqref{sys}, yields the linear dynamics
\begequ
\lab{lindyn}
{\tilde Y}^{(3)} + k_2 \ddot {\tilde Y}  + k_1 \dot {\tilde Y}   + k_0 {\tilde Y}=0,
\endequ  
where $\tilde {(\cdot)}:=(\cdot)-(\cdot)_\star$.

\begin{prop}
\lab{pro5}
Consider the model of the levitated ball system \eqref{sys}. Fix a desired vertical position $Y_\star  \in (-\infty,c)$ with associated equilibrium $x_\star:=\col(Y_\star,0,\sqrt{2kmg})$. Define the sensorless position controller as the certainty equivalent version of \eqref{feelin}
where $\hat \lambda,\hat v,\hat Y$ are generated via the observers of Propositions \ref{pro2}, \ref{pro3} and \ref{pro4}, respectively, and the coefficients $k_i,\;i=0,1,2$, are chosen to ensure that the system \eqref{lindyn} is stable.

Assume $\Delta \not\in \mathcal{L}_2$, $\hat \lambda \not\in \mathcal{L}_2$ and {\bf Assumptions A1, A2} hold.
\begite
\item[(i)] The overall closed-loop dynamics is given by \eqref{dotelam},  \eqref{dotev}, \eqref{dotey} and
$$
\dot {\tilde x} =A \tilde x +\epst,
$$
where $x$ is defined in \eqref{x} and $A$ is a Hurwitz matirx.
\item[(ii)] There exists a (sufficiently small) constat $\delta>0$ such that the following implication holds
\endite
\begalis{
&e^2_\lambda(0)+e^2_Y(0)+e^2_v(0)+e_x^2(0) \leq \delta  \Longrightarrow \\ & \\
&\lim_{t \to \infty} e_\lambda(t)  =\lim_{t \to \infty}e_Y(t)=\lim_{t \to \infty}e_v(t)=\lim_{t \to \infty} \tilde x(t)=0.
}
\end{prop}

\begin{pf}
First, notice that the certainty equivalent version of the control  \eqref{feelin} may be written in the form $u_{\tt CE}=u_{\tt FSF}+\delta_u$,  where $u_{\tt FSF}$ is the full-state controller. Under the standing assumptions, Propositions \ref{pro2}, \ref{pro3} and \ref{pro4} ensure that $\delta_u(t) \to 0$. 

The proof of claim (i) is completed noting that the dynamics \eqref{sys} is linear in $u$. Claim (ii) is established invoking standard arguments used to analyse stability of cascaded systems, {\em e.g.}, Theorem 3.1 of \cite{VID}. 
\end{pf}
%
\section{Simulations}
\lab{sec5}
The $1$-dof Maglev system \eqref{sys} in closed-loop with the sensorless version of the FLC \eqref{feelin} was simulated with the following plant parameters: $m = 0.0844$, $k = 1$, $R = 2.52$, $c = 0.005$. The filters used in DREM were implemented with the gains $\rho=0.01$, $\mu=10$, while the parameters of the FLC were fixed at $k_0=1000$, $k_1=300$, $k_2=30$, which corresponds to a pole location of the ideal closed-loop dynamics of $s_1 = s_2 = s_3 = -10$. For all experiments the default initial conditions are $\lambda(0) = \eta$, with the value of $\eta$ given later, $\psi(0) = 0$, $\hat \lambda(0) = 0$, $Y(0)=-1$,  $\dot{Y}(0) = 0.5$,  $\hat{Y}(0) = 0$, $\hat v(0)=0$, $\hat{\eta}(0)=0.0001$. 

Two reference signals for $Y$ were considered: filtered sum of sinusoids and filtered steps, namely, 
$$
Y^*(t) = \frac{\nu^4}{(p+\nu)^4}Y^*_0(t),
$$
with
\begin{align}
	Y^*_0(t) &= \sin t + \sin 2t + 0.5\sin(3.7t+\pi/3),
\\
\textnormal{and}\qquad\quad&\nonumber
\\
\lab{steps2}
Y^*_0(t)&=\left\{\begin{array}{l}
0, \text{ for } 0\le t\le 1 \text{ sec, }\\
2, \text{ for } 1\le t\le 3 \text{ sec, }\\
0, \text{ for } 3\le t\le 5 \text{ sec, }\\
3, \text{ for } t\ge 5 \text{ sec. }
\end{array}
\right.
\end{align} 
where $\nu=10$ for the sinusoids and $\nu=1$ for the steps. 

\begin{figure}
	\vspace{-6mm}
	\centering
	\includegraphics[width=0.35\textwidth]{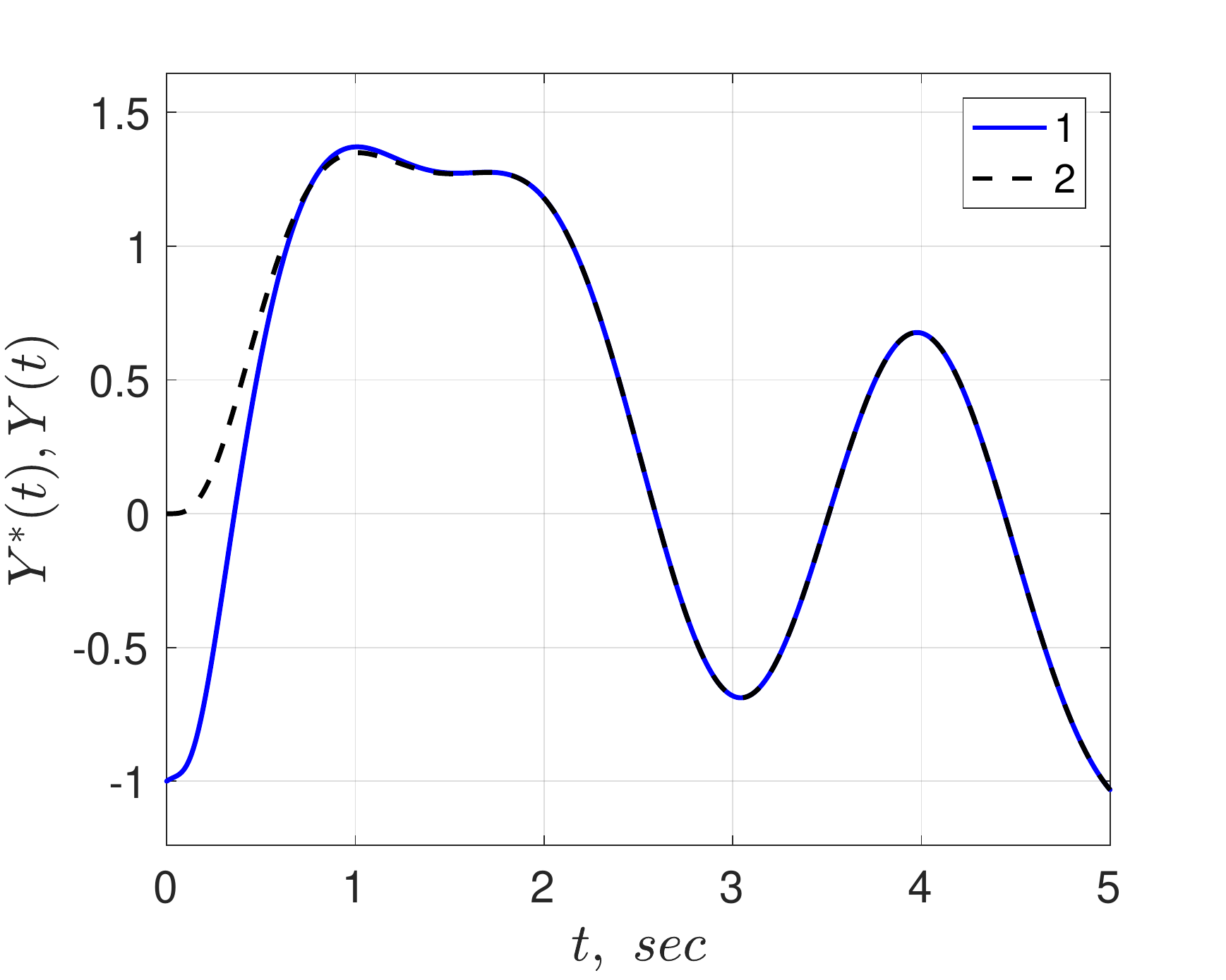}
	\vspace{-3mm}
	\caption{\label{fig:1dof_Y_1} Filtered sum of sinusoids with  $\gamma=1$ and $\eta = 0.01$}
\end{figure}

\begin{figure}	
	\centering
	\includegraphics[width=0.35\textwidth]{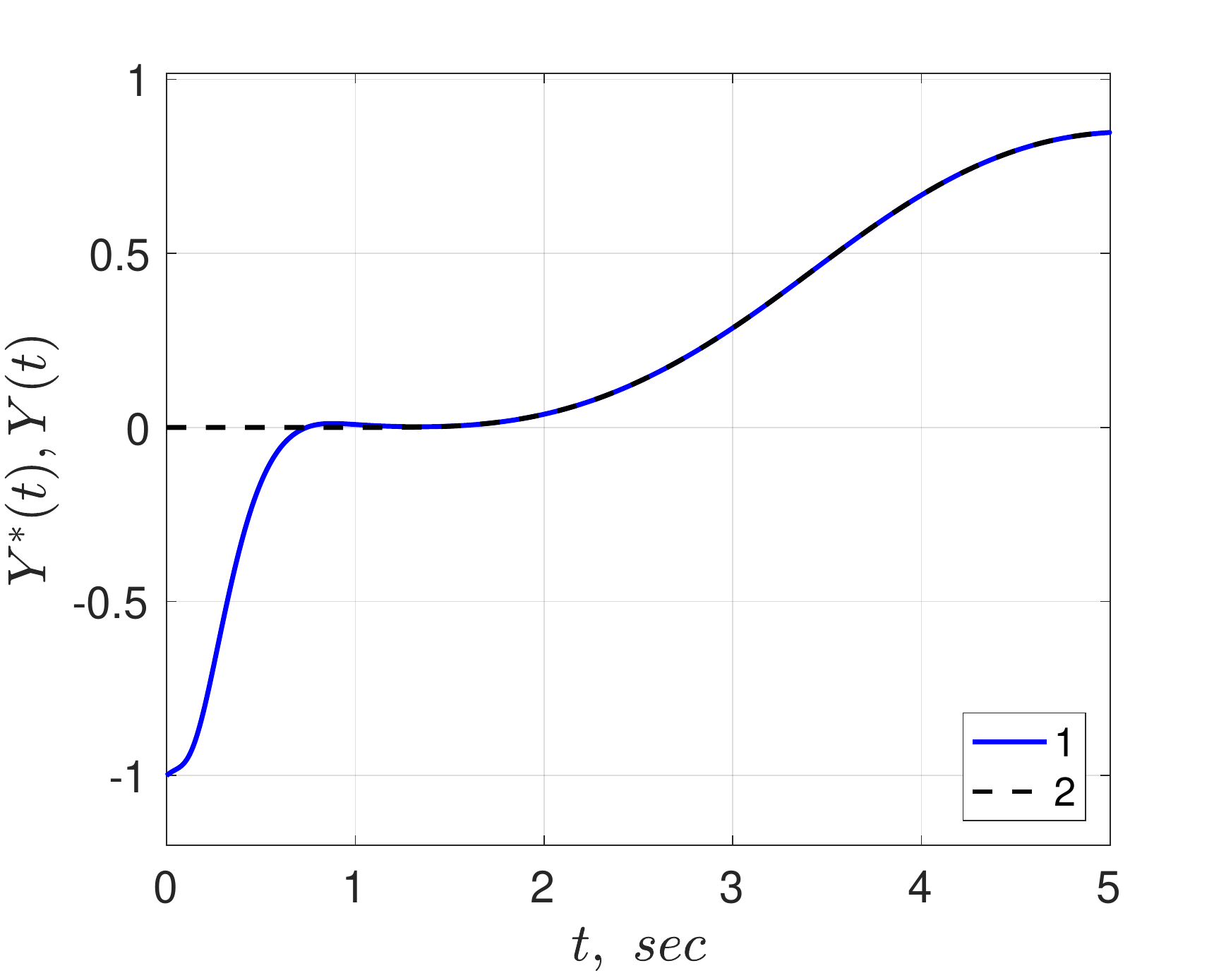}
	\vspace{-3mm}
	\caption{\label{fig:1dof_Y_2} Filtered steps with $\gamma=10^3$ and $\eta = 0.01$}
\end{figure}

\begin{figure*}[htp]
	\centering
	\subcaptionbox{\label{fig:1dof_errors_sin_gamma_lambda} Transients for $\lambda(t)-\hat\lambda(t)$}{\includegraphics[width=0.31\textwidth]{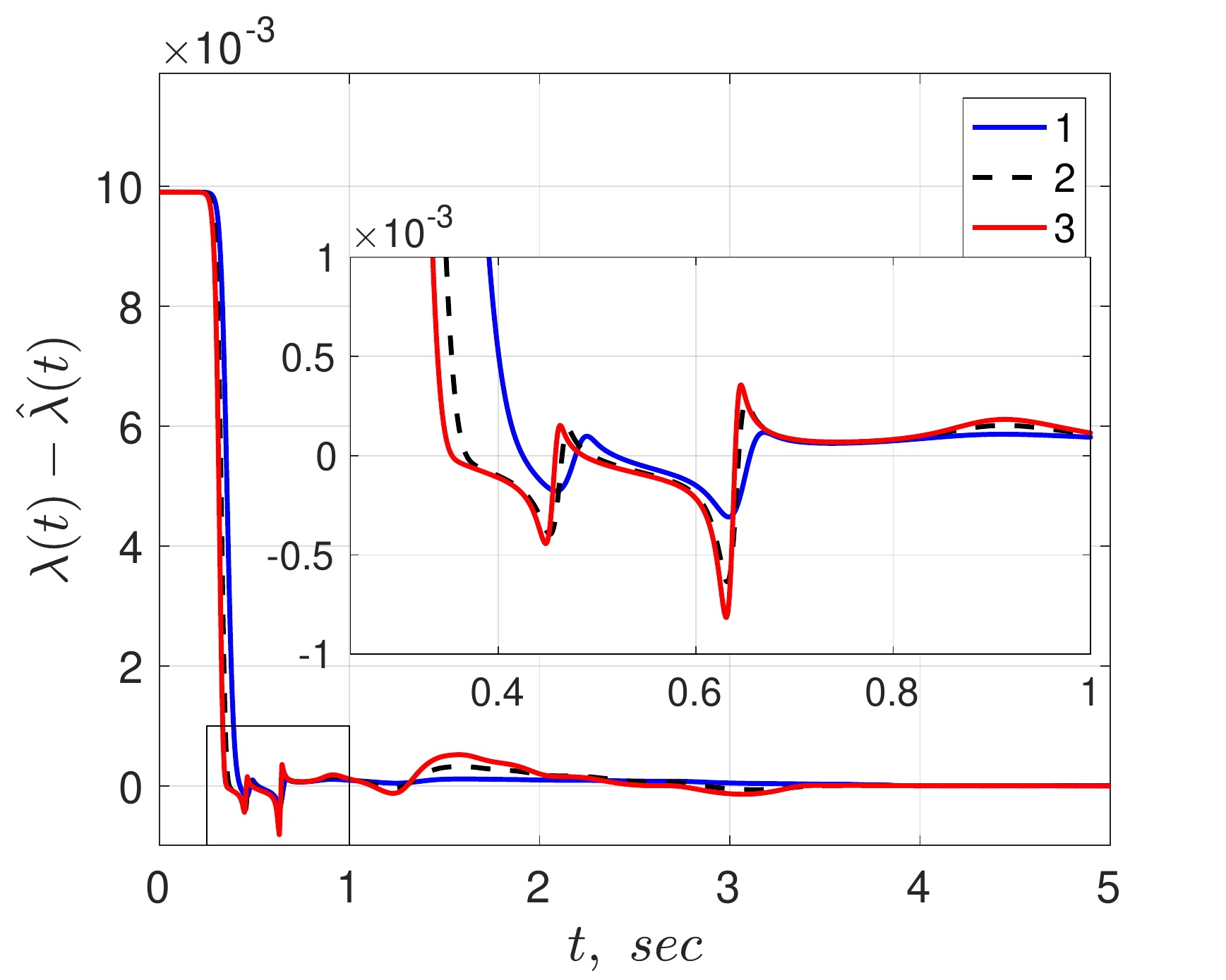}}
	\
	\subcaptionbox{\label{fig:1dof_errors_sin_gamma_Y} Transients for $Y(t)-\hat Y(t)$}{\includegraphics[width=0.31\textwidth]{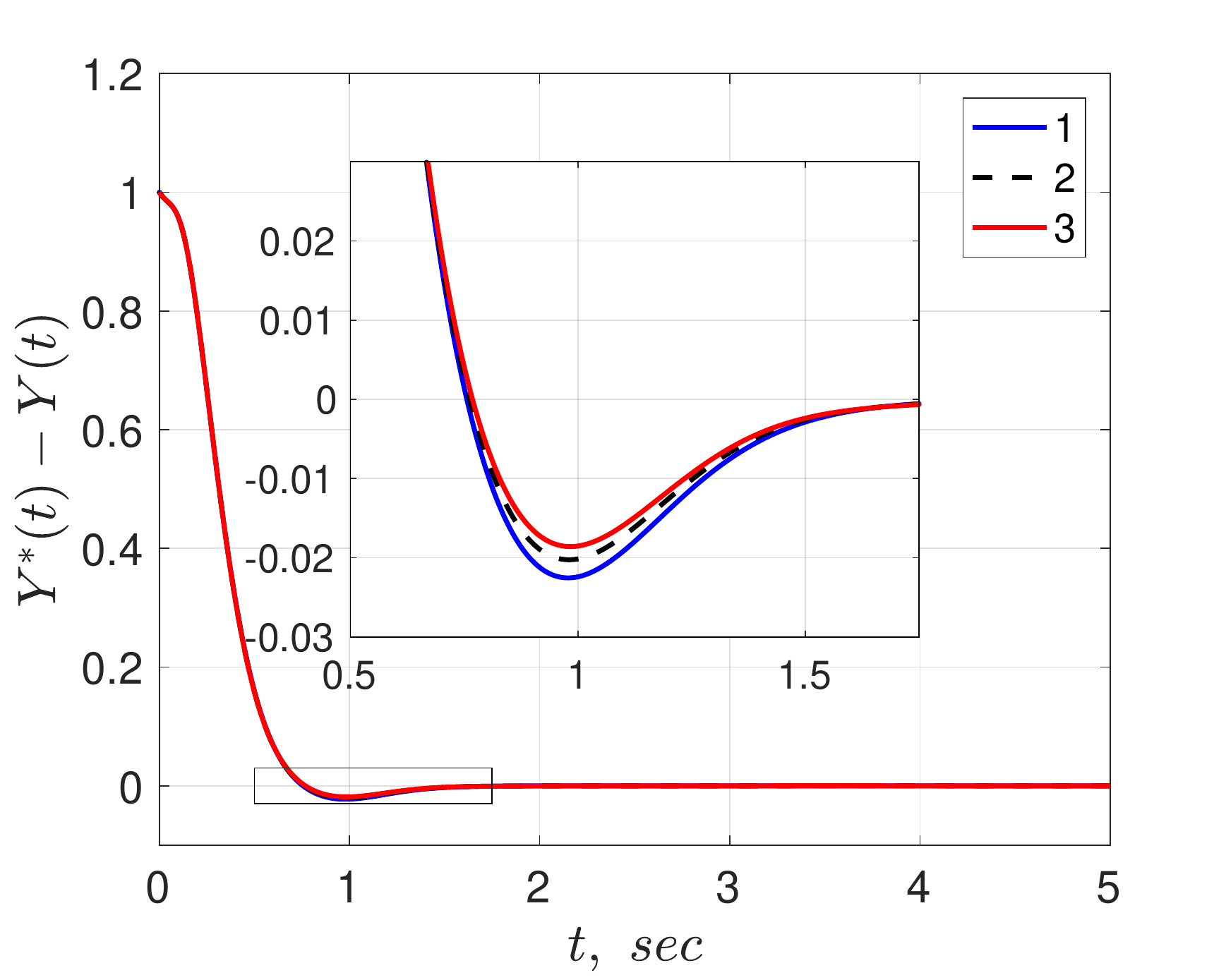}}
	\
	\subcaptionbox{\label{fig:1dof_errors_sin_gamma_dY} Transients for $\dot Y(t)-\hat v_{Y}(t)$}{\includegraphics[width=0.31\textwidth]{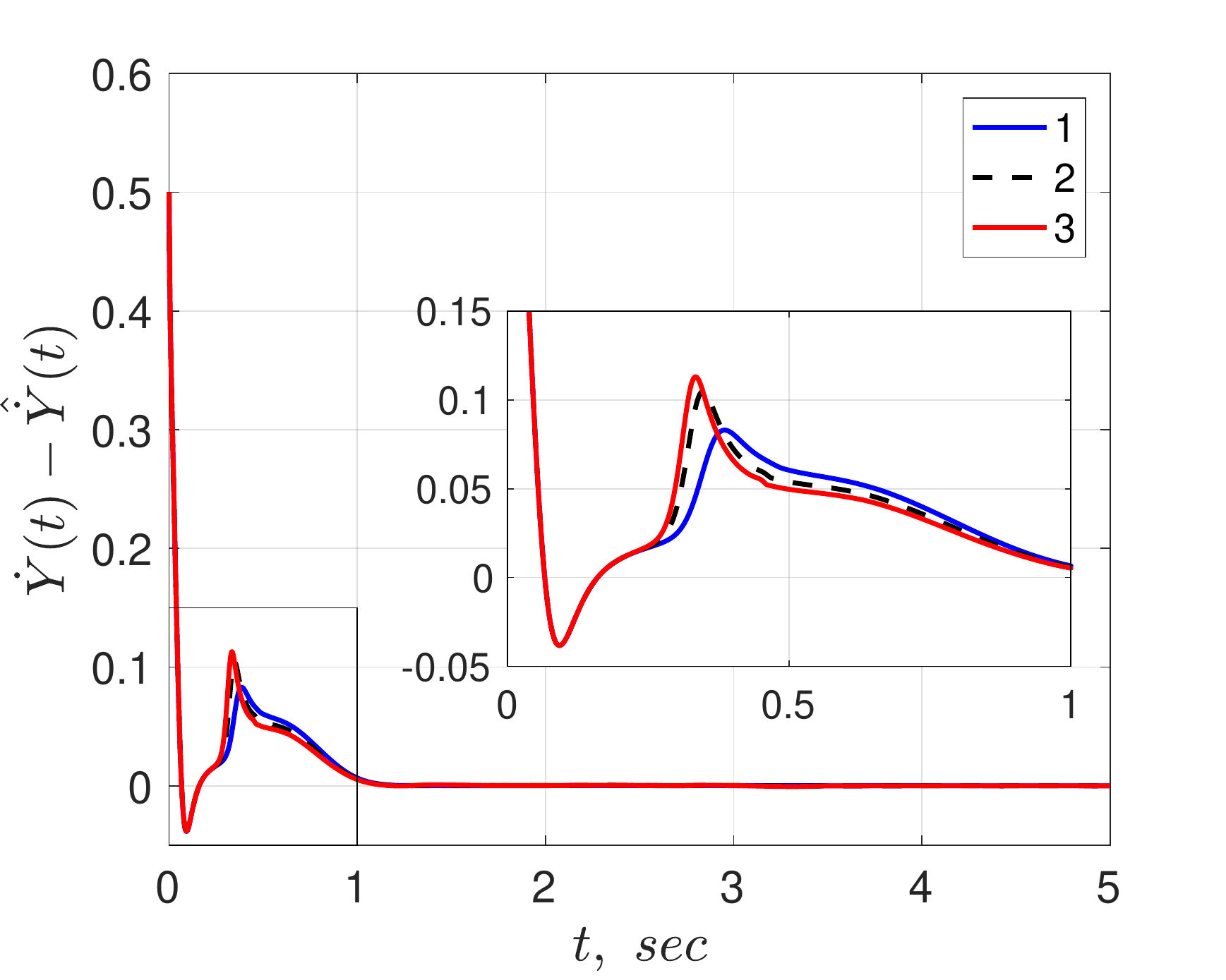}}
	\vspace{-2mm}
	\caption{\label{fig:1dof_errors_sin_gamma} Errors with the sensorless-based FLC for the sinusoidal position reference: 1. $\gamma=1$, 2. $\gamma=5$, 3. $\gamma=10$}
\end{figure*}

\begin{figure*}[htp]
	\centering
	\subcaptionbox{\label{fig:1dof_errors_step_gamma_lambda} Transients for $\lambda(t)-\hat\lambda(t)$}{\includegraphics[width=0.31\textwidth]{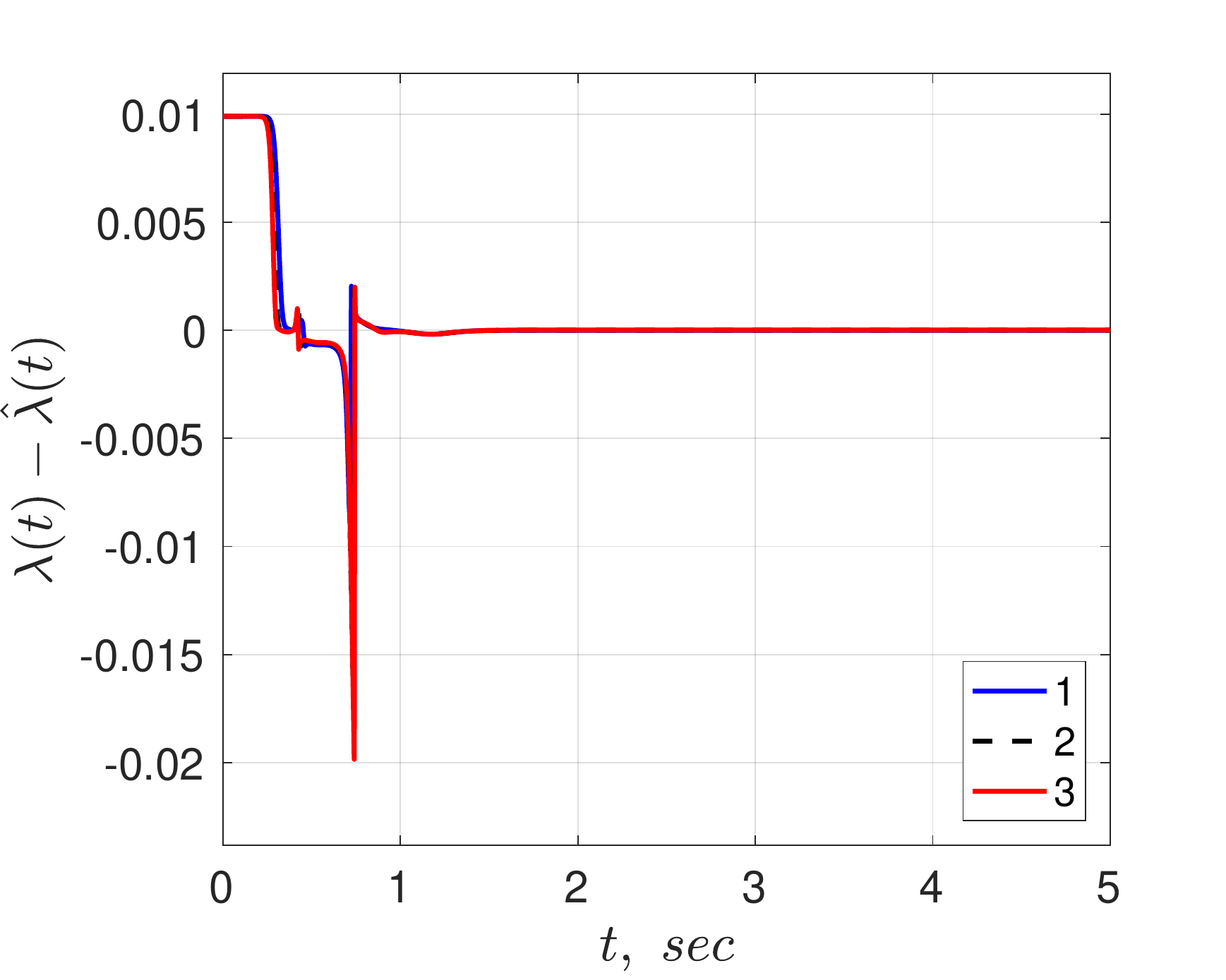}}
	\
	\subcaptionbox{\label{fig:1dof_errors_step_gamma_Y} Transients for $Y(t)-\hat Y(t)$}{\includegraphics[width=0.31\textwidth]{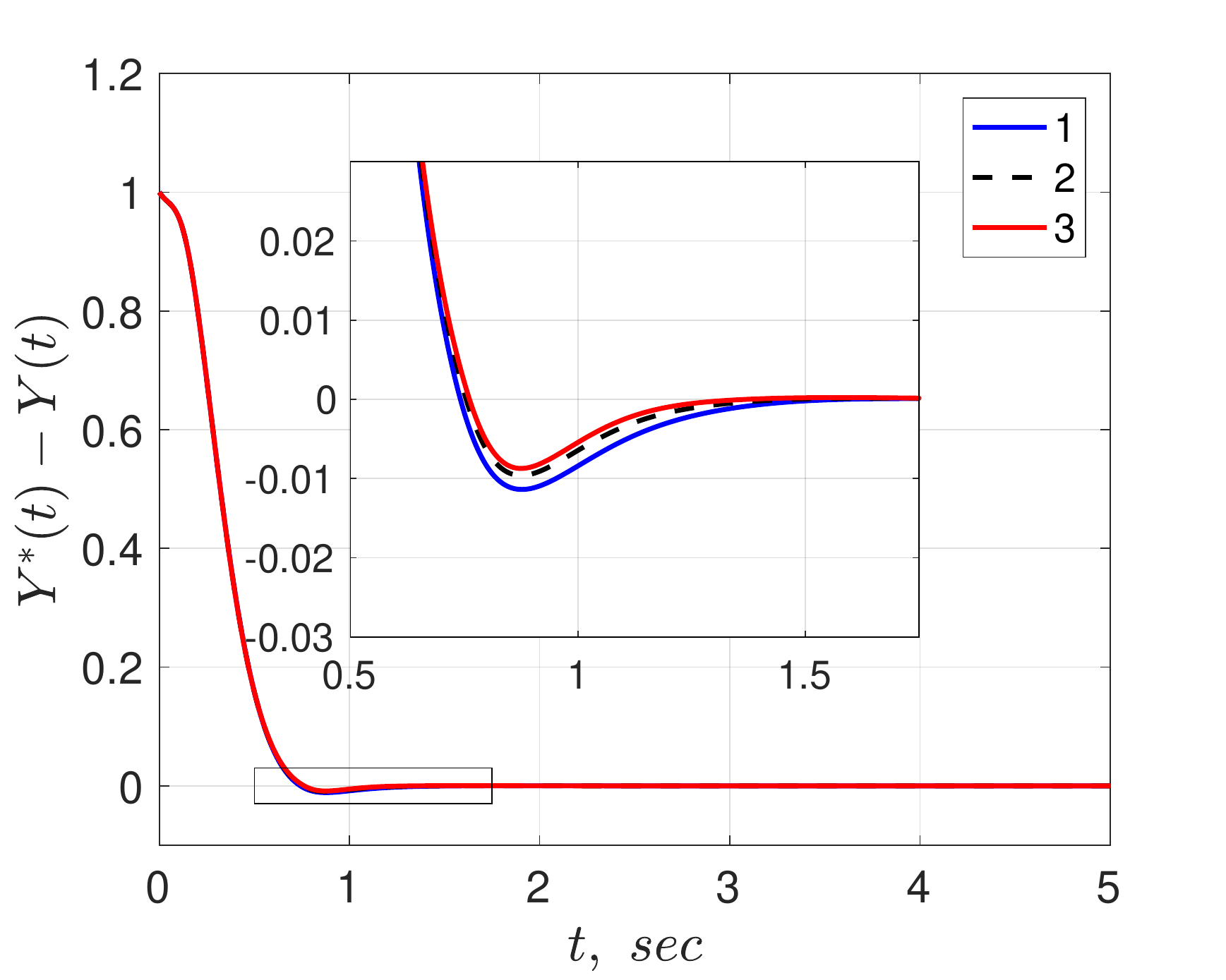}}
	\
	\subcaptionbox{\label{fig:1dof_errors_step_gamma_dY} Transients for $\dot Y(t)-\hat v_{Y}(t)$}{\includegraphics[width=0.31\textwidth]{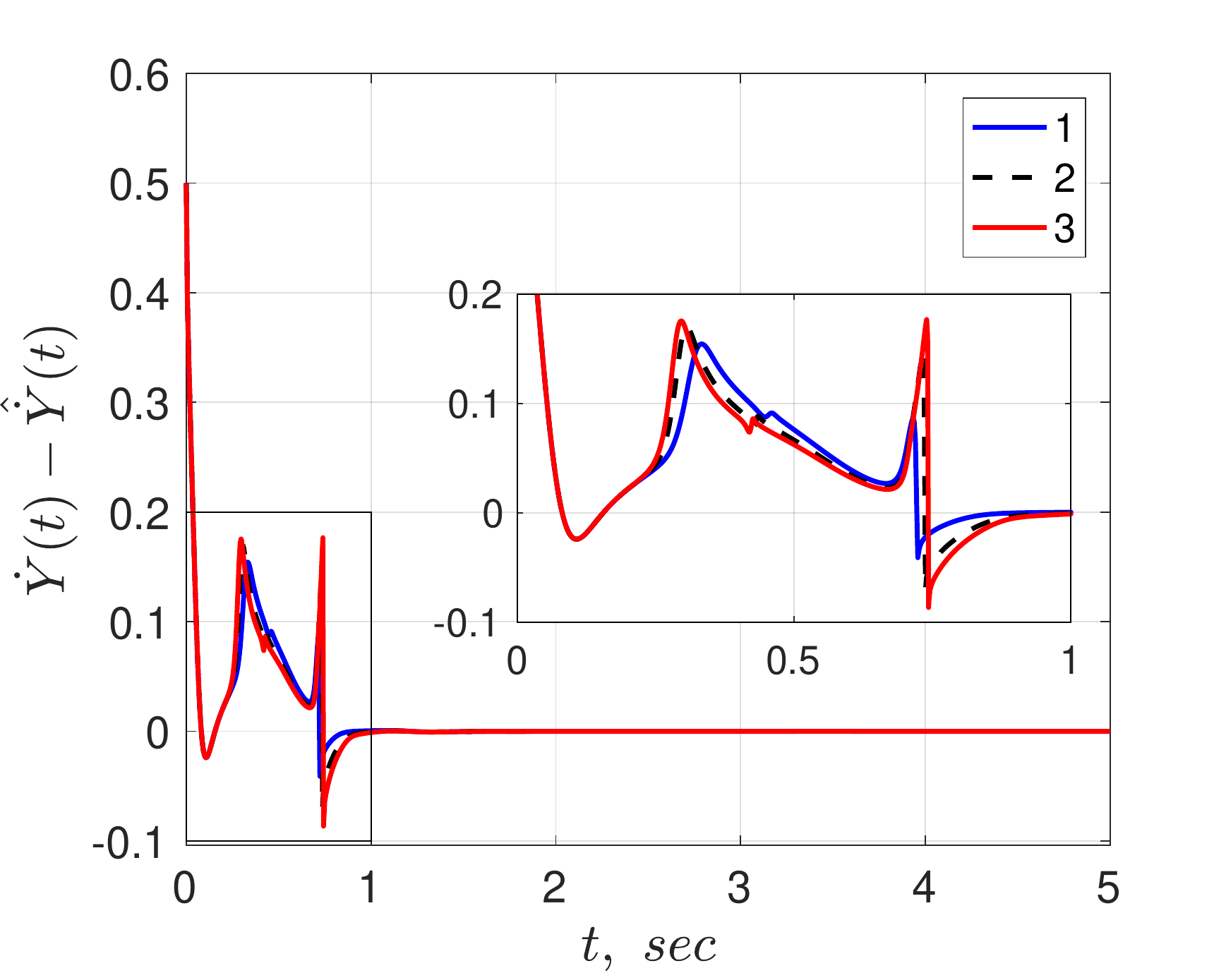}}
	\vspace{-2mm}
	\caption{\label{fig:1dof_errors_step_gamma} Errors with the sensorless-based FLC for the steps position reference: 1. $\gamma=1000$, 2. $\gamma=5000$, 3. $\gamma=10000$}
\end{figure*}

\begin{figure*}[htp]
	\centering
	\subcaptionbox{\label{fig:1dof_errors_sin_l0_lambda} Transients for $\lambda(t)-\hat\lambda(t)$}{\includegraphics[width=0.31\textwidth]{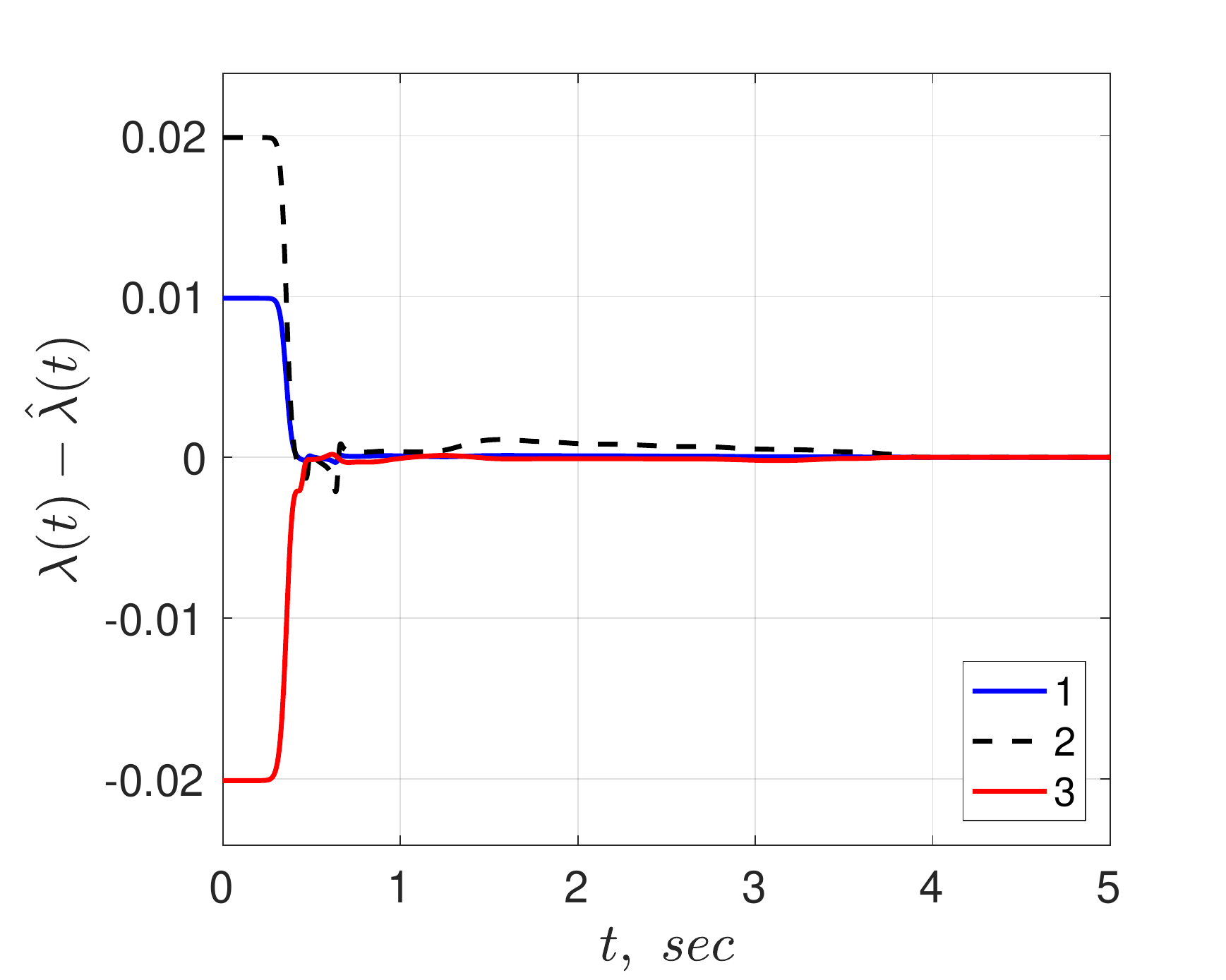}}
	\
	\subcaptionbox{\label{fig:1dof_errors_sin_l0_Y} Transients for $Y(t)-\hat Y(t)$}{\includegraphics[width=0.31\textwidth]{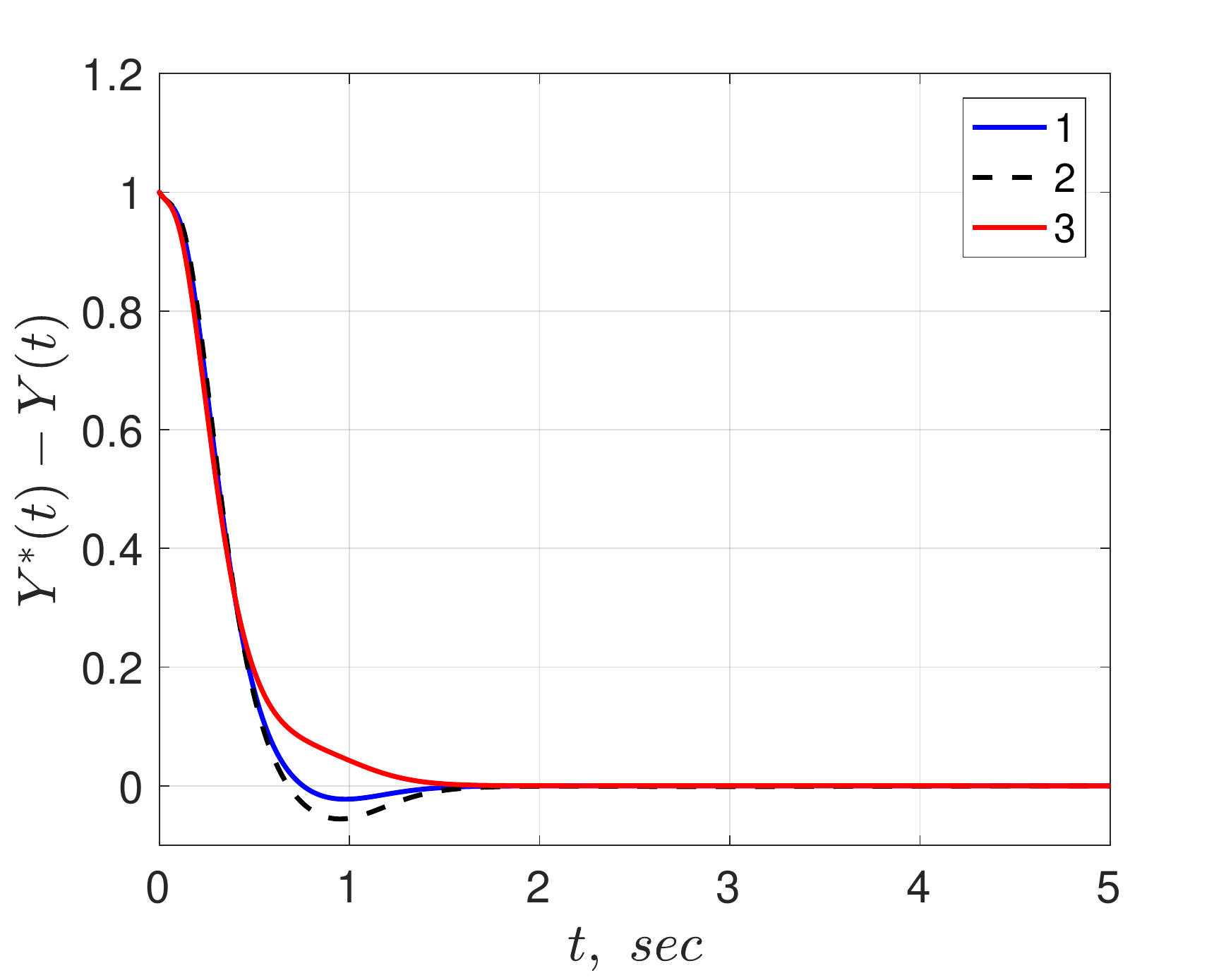}}
	\
	\subcaptionbox{\label{fig:1dof_errors_sin_l0_dY} Transients for $\dot Y(t)-\hat v_{Y}(t)$}{\includegraphics[width=0.31\textwidth]{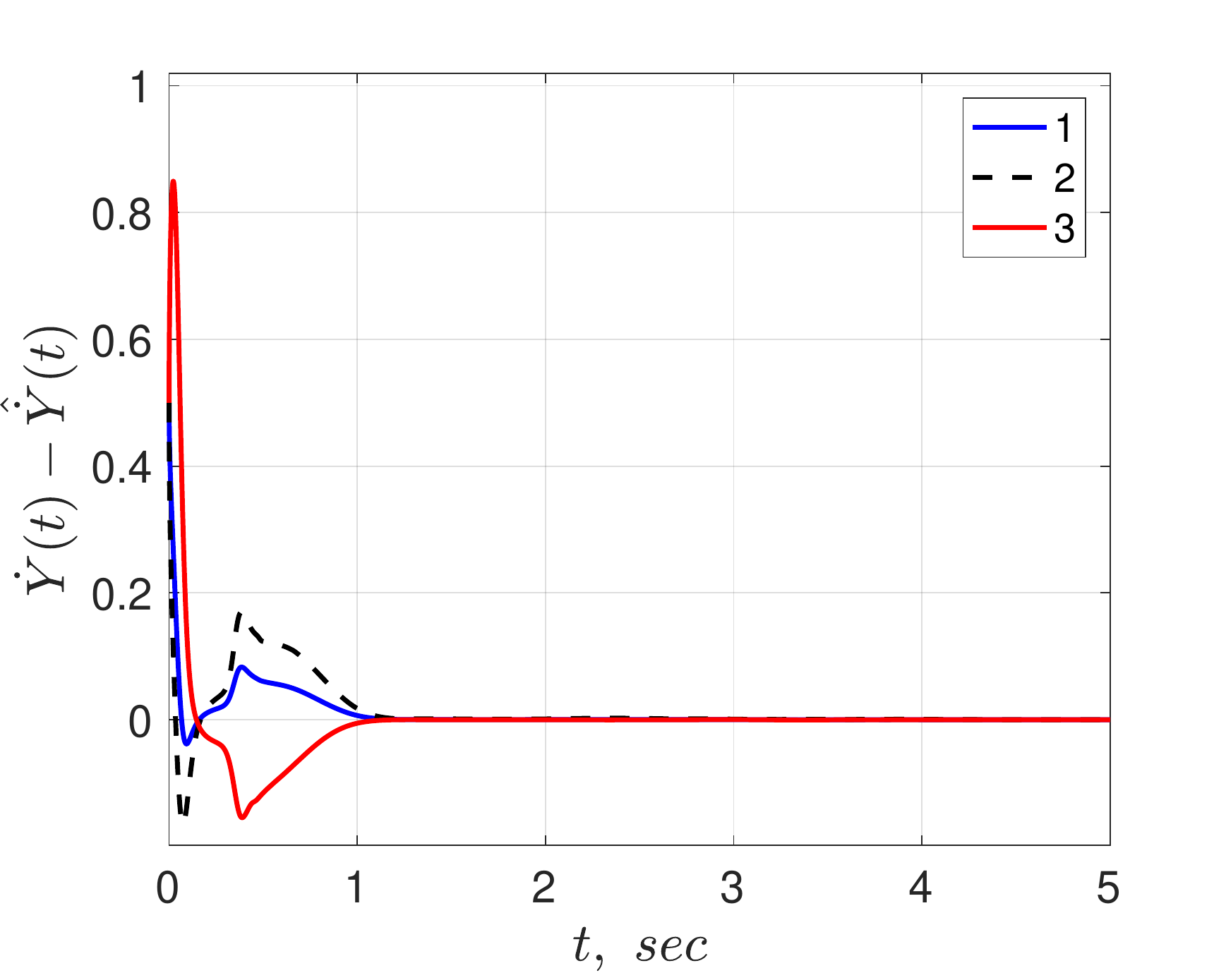}}
	\vspace{-2mm}
	\caption{\label{fig:1dof_errors_sin_l0} Errors with the sensorless-based FLC for the sinusoidal position reference: 1. $\eta=0.01$, 2. $\eta=0.02$, 3. $\eta=-0.02$}
\end{figure*}

\begin{figure*}[htp]
	\centering
	\subcaptionbox{\label{fig:1dof_errors_step_l0_lambda} Transients for $\lambda(t)-\hat\lambda(t)$}{\includegraphics[width=0.31\textwidth]{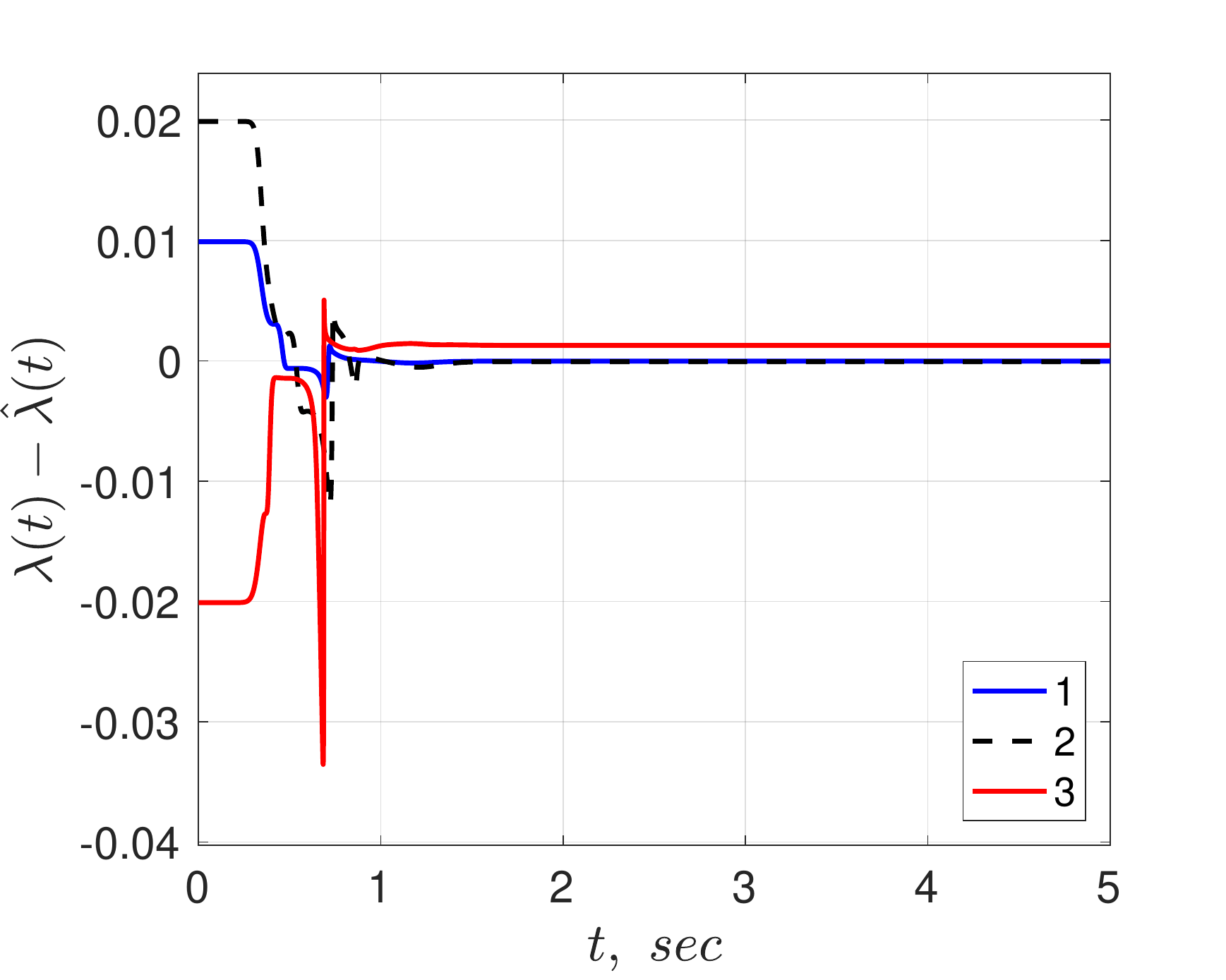}}
	\
	\subcaptionbox{\label{fig:1dof_errors_step_l0_Y} Transients for $Y(t)-\hat Y(t)$}{\includegraphics[width=0.31\textwidth]{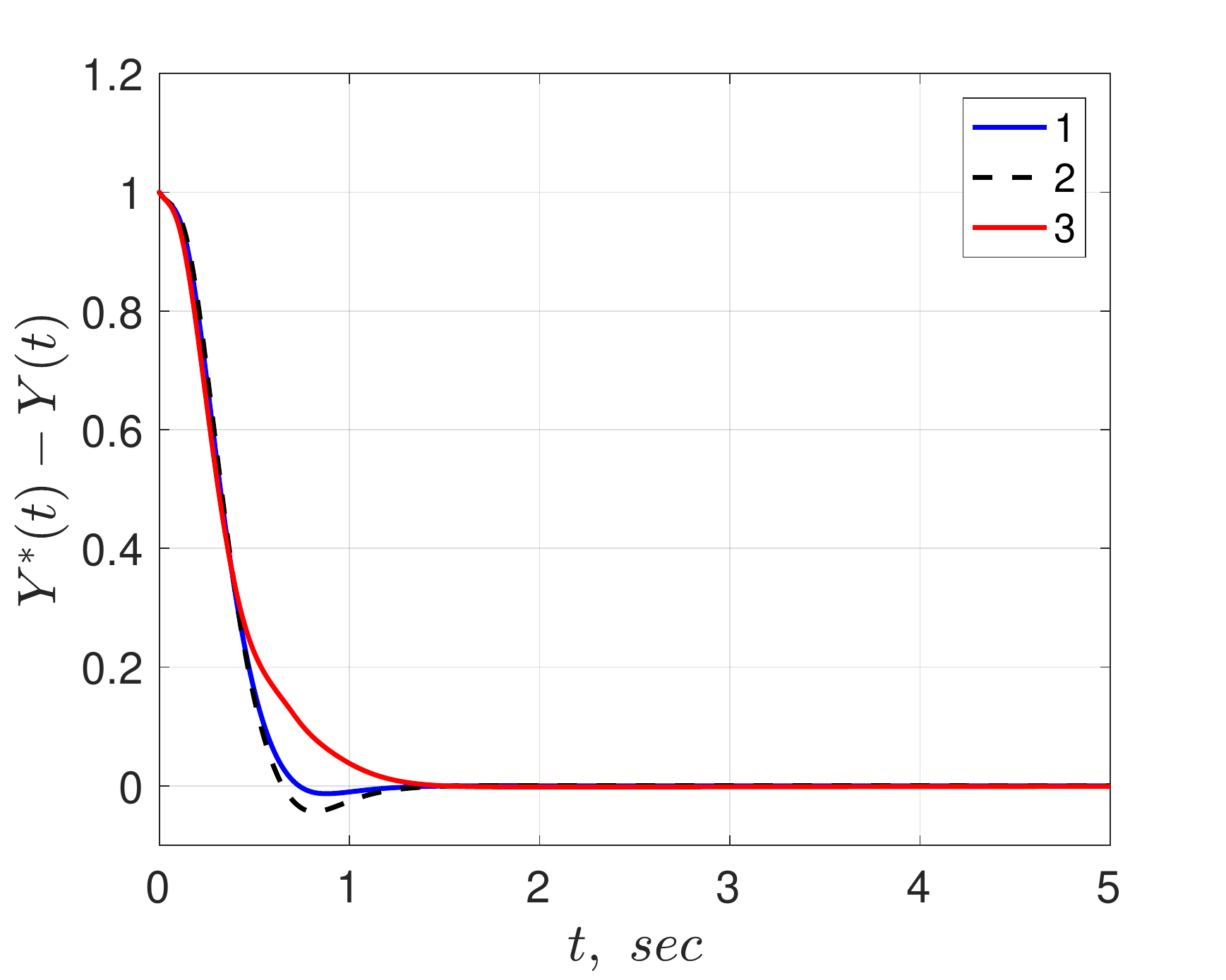}}
	\
	\subcaptionbox{\label{fig:1dof_errors_step_l0_dY} Transients for $\dot Y(t)-\hat v_{Y}(t)$}{\includegraphics[width=0.31\textwidth]{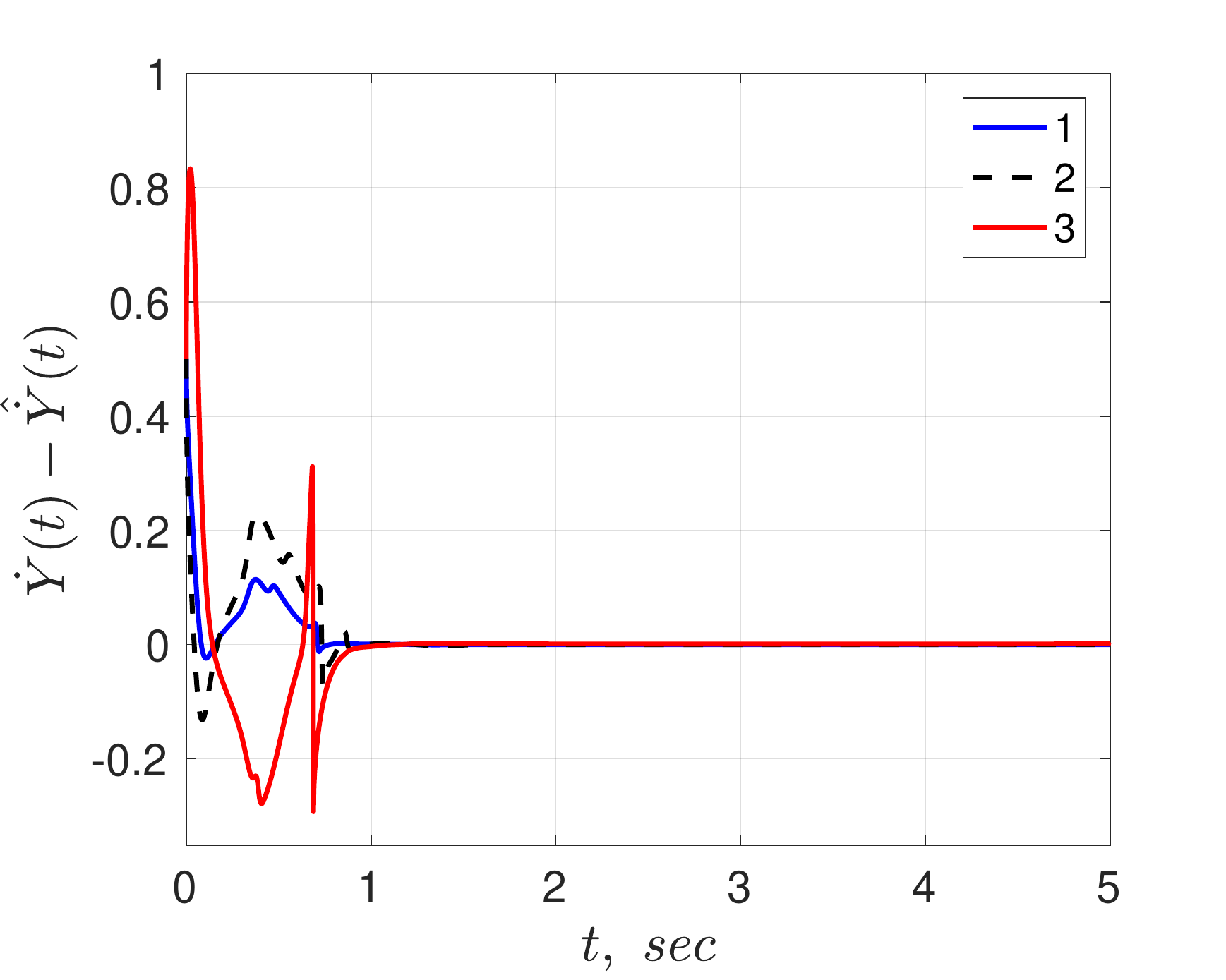}}
	\vspace{-2mm}
	\caption{\label{fig:1dof_errors_step_l0} Errors with the sensorless-based FLC for the smooth steps position reference: 1. $\eta=0.01$, 2. $\eta=0.02$, 3. $\eta=-0.02$}
\end{figure*}

In Figs. \ref{fig:1dof_Y_1} and~\ref{fig:1dof_Y_2} we compare the behaviour of the position for the two desired trajectories with the difference in the initial conditions of $\lambda$ and $\psi$ such that $\eta = 0.01$, $\lambda(0) = 0.01$ and $\psi(0) = 0$.
In Figs.~\ref{fig:1dof_errors_sin_gamma} and  \ref{fig:1dof_errors_step_gamma} we evaluated the effect on the observation errors of changing the flux observer adaptation gain $\gamma$. In Figs. \ref{fig:1dof_errors_sin_l0} and \ref{fig:1dof_errors_step_l0} the behaviour of the observer for different values of $\eta$ is showed. In last figure we observe that there is a steady state error, which increases for bigger adaptation gains. This reveals that the condition $\Delta \notin \call_2$ is not satisfied, but the overall performance is still satisfactory.



\section{Conclusions and Future Research}
\lab{sec6}
\vspace{-2mm}
%
We have presented in this paper the first solution to the challenging problem of designing a sensorless controller for the levitated ball system, without signal injections. Instrumental for the development of the theory was the use of PEBO and DREM parameter estimators---which were recently reported in the control literature---to estimate the flux and the mechanical coordinates of the system. The sensorless controller is then obtained replacing the estimated state in a full-state feedback FLC. It should be underscored that these controller can be replaced with any other full-state feedback stabilizing controller. Simulation results show the excellent behaviour of the proposed observer. Consequently, the regulation performance of the sensorless controller is very similar to the one obtained with the full-state feedback scheme. 

The convergence proof of the proposed observers relies on excitation conditions that are hard to verify {\em a-priori}. Moreover, these conditions are critically dependent on the choice of the filters that generate the extended regressors~---~see \cite{ARAetaltac,ORTetalaut} for some discussion on this important issue.  

Several open questions are currently being investigated.  
\bul The computational complexity of the proposed observer is relatively high for this kind of application. Controller approximation techniques should be tried to obtain a practical design. 

\bul Experimental validation is currently under way, but is being hampered by the computational complexity mentioned above.

\bul It would be interested to compare our proposal with existing technique-oriented methods as well as the signal injection-based PEBO reported in \cite{yicdc18}---where an experimental validation was already carried out.

\bul Saturation effects, which may degrade the systems performance, have been neglected in our analysis. It seems possible to incorporate this consideration in the controller design.

\bul As mentioned in Remark \ref{rem20} a potential difficulty of DREM is the use of open-loop integration. This problem is particularly important in for noisy signals. It should be mentioned that, in spite of this potential drawback, several successful experimental validations of the effectiveness of PEBO, which incorporate some {\em ad-hoc} ``safety-nets" to PEBO, have been reported, see {\em e.g.}, \cite{BOBetalijc,CHOetal}. Finding the right safety nets for the MagLev application will be needed in the experimental test. 

\bibliography{references}

%
\appendix 
%
\section{Proof of Proposition \ref{pro5}}
\lab{propro5}
%
To simplify the expressions we write the model \eqref{sys} in state-space form with the state variables $x=\col(x_1, x_2, x_3):=\col(Y, m\dot Y, \lambda)$ and denote the measurable signal $y:=i$. This yields,
\begin{align}
\lab{x1}
\dot x_1&={1\over m}x_2,\\
\lab{x2}
\dot x_2 &={1\over 2k}x_3^2-mg,\\
\lab{x3} 
\dot x_3 & = -Ry+u,\\
\lab{1DOF_y}
y & = {1\over k}(c-x_1)x_3.
\end{align}

From \eqref{dotpsi0} and \eqref{x3} we get
\begin{align}
\lab{x3_par}
x_3(t)=\eta+\psi(t),
\end{align}
where $\eta=x_3(0)-\psi(0)$. The essence of the proof is to,  using \eqref{x3_par}, manipulate the systems equations \eqref{x1}--\eqref{1DOF_y} to establish an algebraic relation that depends only on the signals $y$ and $u$ ---and filtered combinations of them---and a function of the unknown parameter $\eta$. 

Instrumental to carry out this task is the Swapping Lemma, see {\em e.g.}, Lemma 3.6.5 of \cite{SASBOD}, that is used in this proof in the following way
$$
{\mu\over p+\mu}[a(t)b(t)]=b(t) {\mu\over p+\mu}[a(t)]+ {\mu\over p+\mu}[\dot b(t){1 \over p+\mu}[a(t)]],
$$
where $a$ and $b$ are some scalar functions of time and $\mu>0$. 

First, compute $\dot y$
\begin{align}
\lab{ydot}
\dot y & = -{1\over k}\dot x_1x_3+{1\over k}(c-x_1)\dot x_3
\end{align}
and consider $y\dot x_3-\dot y x_3$ together with \eqref{x1}:
\begin{align}
\lab{reg1}
y\dot x_3-\dot y x_3 & = {1\over k}\dot x_1x_3^2=
{1\over km} x_2x_3^2.
\end{align}
Substituting \eqref{x3_par} into \eqref{reg1} we get
\begin{align}
\lab{reg3}
y\dot x_3-\dot y \psi &=\dot y \eta+
{1\over km} x_2x_3^2.
\end{align}
Applying the operator\footnote{To simplify the notation, In the sequel we omit the argument $p$ from the operator $W(p)$.}
$$
W(p):={\mu\over p+\mu}.
$$
to \eqref{reg3} we get
\begin{align}
\lab{reg4}
W [y\dot x_3-\dot y \psi] &=W \dot y \eta+
W {1\over km} x_2x_3^2.
\end{align}
Define the signal
\begin{align}
r_1 :& =  W \,y\dot x_3-W \,\dot y \psi \nonumber\\
	& = W\, y(u-Ry)-\psi\,W \,\dot y+ {1\over p+\mu}\,\left[\dot\psi W \,\dot y\right]\nonumber\\
	& = W\, y(u-Ry)-\psi\,{\mu\, p\over p+\mu} \, y \nonumber \\
	& \quad + {1\over p+\mu}\,\left[(u-Ry) {\mu\, p\over p+\mu} \, y\right]\nonumber\\
	& = W\, y(u\!-\!Ry)\!-\!\psi\,\omega_1 \! +\! {1\over p\!+\!\mu}\,\left[(u\!-\!Ry) \omega_1\right],
\lab{q1}
\end{align}
where the Swapping Lemma  was applied to the term $W \,\dot y \psi$ to get the second identity and we defined the (measurable) signal 
\begin{align}
\lab{omega1}
\omega_1:=W \, y.
\end{align}
Note that $r_1$ may be computed based on $y$ and $u$ only.
Replacing \eqref{q1} in \eqref{reg4} we get
\begin{align}
\lab{reg5}
km\,r_1 &= \eta km \omega_1
+
W x_2x_3^2
\end{align}
and after applying the Swapping Lemma again to the term $W x_2x_3^2$ we get
\begin{align}
km\,r_1 & = \eta \,km\,\omega_1 + x_2W x_3^2-{1\over p+\mu}\left[\dot x_2W x_3^2\right] \nonumber \\
	& = \eta \,km\,\omega_1 + x_2W x_3^2 \nonumber \\
	& \quad - {1\over p+\mu}\left[\left({1\over 2k}x_3^2-mg\right)W x_3^2\right] \nonumber \\
	& = \eta \,km\,\omega_1 + x_2\phi_1 \nonumber \\
	& \quad - {1\over p+\mu}\left[\left({1\over 2k}x_3^2-mg\right)\phi_1\right]
\lab{reg6}
\end{align}
where we defined the signal
\begin{align}
\lab{phi1}
\phi_1:=W x_3^2.
\end{align}
Define a second auxiliary signal
\begin{align}
\lab{q2}
r_2 :& = W \,km\,r_1 \nonumber \\
	& = \eta \,km\,W\,\omega_1 + {W\,x_2\phi_1} \nonumber \\
	& \quad - {\mu\over (p+\mu)^2}\left[\left({1\over 2k}x_3^2-mg\right)\phi_1\right] \nonumber \\
	& = \eta \,km\,W\,\omega_1 + {x_2W\,\phi_1-{1\over p+\mu}\left[\dot x_2W\,\phi_1\right]} \nonumber \\
	& \quad - {\mu\over (p+\mu)^2}\left[\left({1\over 2k}x_3^2-mg\right)\phi_1\right] \nonumber \\
	& = \eta \,km\,\omega_2 + x_2\phi_2 - {1\over p+\mu}\left[\left({1\over 2k}x_3^2-mg\right)\phi_2\right] \nonumber \\
	& \quad - {\mu\over (p+\mu)^2}\left[\left({1\over 2k}x_3^2-mg \right)\phi_1\right]
\end{align}
where we used \eqref{reg6} in the second equation, applied the Swapping Lemma to the term $W x_2\phi_1$ to get the third identity, used \eqref{x2} in the fourth one and  
\begin{align}
	\omega_2:=W\,\omega_1,
	\\
	\phi_2:=W\,\phi_1.
	\lab{phi2}
\end{align}
Consider the following identity
\begin{align}
	& (km\,r_1\phi_2 - r_2\phi_1)2k\mu = \eta\,(\omega_1\phi_2-\omega_2\phi_1)2k\mu \nonumber \\
	& \quad - \phi_2\,W \left[x_3^2\phi_1-2mgk\phi_1\right] + \phi_1\,W\left[x_3^2\phi_2-2mgk\phi_2\right] \nonumber \\
	& \quad	+ \phi_1\,{\mu\over (p+\mu)^2}\left[x_3^2\phi_1-2mgk\phi_1\right],
\lab{eq:q3}
\end{align}
where we replaced $r_1$ and $r_2$ with \eqref{reg6} and~\eqref{q2} respectively to obtain right-hand side. 
Signals $x_3^2$, $\phi_1$, and $\phi_2$ cannot be computed based on the measurable signals $y$ and $u$, but can be replaced by combination of the measurable signal $\psi$ and unknown parameter $\eta$ using ~\eqref{x3_par}, \eqref{phi1}, and~\eqref{phi2}
\begin{align}
\lab{x32}
x_3^2 &= \eta^2 + 2\eta\psi + \psi^2, \\
\lab{phi1_par}
\phi_1 &= \eta^2 + 2\eta(W\psi) + (W\psi^2) + \et,\\
\lab{phi2_par}
\phi_2 &= \eta^2 + 2\eta(W^2\psi) + (W^2\psi^2) + \et.
\end{align}
Neglecting the exponential decaying terms $\et$ in~\eqref{phi1_par}--\eqref{phi2_par} and substituting with~\eqref{x32} into~\eqref{eq:q3} after lenghty, but straightfoward, calculations we get a linear regression model:
\begin{align}
\lab{eq:1dof_reg}
z^0 = \eta\varphi^0_1 + \eta^2\varphi^0_2
  + \eta^3\varphi^0_3 + \eta^4\varphi^0_4
  + \eta^5\varphi^0_5 + \eta^6,
\end{align}
where
\break
\vspace{-3mm}
\begin{align}
z^0 & := 2 k \mu ( k m r_1 \phi_2  -  r_2 \phi_1 ) \nonumber \\
  & \quad - W[\psi^2] \Big( W[(\psi^2 - 2 m g k) W^2[\psi^2]] \nonumber \\
  & \quad \quad + W^2[(\psi^2 - 2 m g k) W[\psi^2]] \Big) \nonumber \\
  & \quad + W^2[\psi^2] W[(\psi^2 - 2 m g k) W[\psi^2]],
\nonumber \\
\varphi^0_1  & := 2 k^2 m \mu ( \omega_1 W^2[\psi^2] - \omega_2 W[\psi^2] ) \nonumber \\
  & \quad + 2 W[\psi]\Big(W[(\psi^2 - 2 m g k) W^2[\psi^2]]  \nonumber \\
  & \quad \quad + W^2[(\psi^2 - 2 m g k) W[\psi^2]]\Big) \nonumber \\
  & \quad + 2 W[\psi^2]\Big( W[\psi W^2[\psi^2]] + W[(\psi^2 - 2 m g k) W^2[\psi]] \nonumber \\
  & \quad \quad + W^2[\psi W[\psi^2]] + W^2[(\psi^2 - 2 m g k)W[\psi]] \Big) \nonumber \\
  & \quad - 2 W^2[\psi] W[(\psi^2 - 2 m g k) W[\psi^2]] \nonumber \\
  & \quad - 2 W^2[\psi^2] ( W[\psi W[\psi^2]] + W[(\psi^2 - 2 m g k) W[\psi]] ),
\nonumber 
\\
\varphi^0_2 & := 4 k^2 m \mu ( \omega_1 W^2[\psi] - \omega_2 W[\psi] ) + (W[\psi_2])^2 \nonumber \\
  & \quad + 2 m g k ( W^2[\psi_2] - 2 W[\psi_2])  \nonumber \\
  & \quad + 4 W[\psi] \Big( W[\psi W^2[\psi_2]] +  W[(\psi^2 - 2 m g k) W^2[\psi]] \nonumber \\
  & \quad \quad	+  W^2[\psi W[\psi_2]] +  W^2[(\psi^2 - 2 m g k) W[\psi]]\Big) \nonumber \\
  & \quad + 2 W[\psi_2] W\Big[W^2[\psi_2] + 2 \psi W^2[\psi] + 2 W [\psi W[\psi]]\Big] \nonumber \\
  & \quad - 4 W^2 \psi ( W[\psi W[\psi_2]] +  W[(\psi^2 - 2 m g k) W[\psi]]) \nonumber \\
  & \quad - (W^2[\psi_2])^2 +  W[(\psi^2 - 2 m g k) W^2[\psi_2]] \nonumber \\
  & \quad + W^2[(\psi^2 - 2 m g k) W[\psi_2]] - 4 W^2[\psi_2] W[\psi W[\psi]] \nonumber \\
  & \quad - W[(\psi^2 - 2 m g k) W[\psi_2]],
\nonumber 
\\
\varphi^0_3 & := 2 k^2 m \mu ( \omega_1 - \omega_2 ) + 4 m g k (W^2[\psi] - 2 W[\psi]) \nonumber \\
  & \quad + 4 W[\psi] \Big( W[\psi^2] + W^3[\psi^2] + 2 W[\psi W^2[\psi]] \nonumber \\
  & \quad \quad + 2 W^2[\psi W[\psi]] \Big) + 4 W[\psi^2] W^3[\psi] \nonumber \\
  & \quad - 4 W^2[\psi] ( W^2[\psi^2] + 2 W[\psi W[\psi]] ) \nonumber \\
  & \quad - 2 W[\psi W[\psi^2]] + 2 W[\psi W^2[\psi^2]] + \nonumber
\\
  & \quad + 2 W[(\psi^2 - 2 m g k)(W^2[\psi] - W[\psi]])] \nonumber \\
  & \quad + 2 W^2[(\psi^2 - 2 m g k) W[\psi] + \psi W[\psi^2]],
\nonumber \\
\varphi^0_4 & := - 2 m g k + 2 W [\psi^2] - W^2[\psi^2] + 2 W^3[\psi^2] \nonumber \\
  & \quad - 4 W[\psi(W[\psi]] + W^2 [\psi]])] + 4 W[\psi] W^2[\psi]  \nonumber \\
  & \quad - 4 (W^2[\psi])^2 + 4 W[\psi] \Big( W[\psi] + 8 W^3[\psi] \Big),
\nonumber \\
\varphi^0_5 & := 4 W[\psi] - 2 W^2[\psi] + 4 W^3 [\psi].
\nonumber
\end{align}

The proof is completed applying to the regression model \eqref{eq:1dof_reg} the filter ${\rho p\over p+\rho}$ to get the new regression model \eqref{reg9}, where we defined
\begalis{
z&:={\rho p\over p+\rho}z^0\\
\phi_i&:={\rho p\over p+\rho}\phi_i^0,\;i=1,\dots,5,
}
 and
$$
\phi:=\col(\phi_1,\dots,\phi_5).
$$ 
Notice that, due to the derivative action of the filter, the constant term $\eta^6$ in \eqref{eq:1dof_reg} has been removed in \eqref{reg9}. This eliminates a constant term (a one) from the regressor, whose excitation conditions for parameter convergence are strictly weaker.

\end{document}